\documentclass[11pt]{article}
\usepackage{amsmath,amsfonts}
\usepackage{bbold}
\usepackage{graphicx,color,multicol,multirow}
\allowdisplaybreaks
\def \beq  {\begin{equation}}
\def \eeq  {\end{equation}}
\def \beqar {\begin{eqnarray}}
\def \eeqar {\end{eqnarray}}
\def\sqr#1#2{{\vcenter{\vbox{\hrule height.#2pt
\hbox{\vrule width.#2pt height#1pt \kern#1pt
\vrule width.#2pt}\hrule height.#2pt}}}}

\def\Tr {{\rm Tr}}

\def\del {\partial}

\def\half{\textstyle{1\over 2}}

\usepackage{enumerate}

\newcommand{\Feynp}[1]{#1\kern-0.42em/}

\def\half{\frac{1}{2}}

\begin{document}
\fontfamily{cmr}\fontsize{11pt}{15pt}\selectfont
\def \CMP {{Commun. Math. Phys.}}
\def \PRL {{Phys. Rev. Lett.}}
\def \PL {{Phys. Lett.}}
\def \NPBProc {{Nucl. Phys. B (Proc. Suppl.)}}
\def \NP {{Nucl. Phys.}}
\def \RMP {{Rev. Mod. Phys.}}
\def \JGP {{J. Geom. Phys.}}
\def \CQG {{Class. Quant. Grav.}}
\def \MPL {{Mod. Phys. Lett.}}
\def \IJMP {{ Int. J. Mod. Phys.}}
\def \JHEP {{JHEP}}
\def \PR {{Phys. Rev.}}
\def \JMP {{J. Math. Phys.}}
\def \GRG{{Gen. Rel. Grav.}}
\begin{titlepage}
\null\vspace{-62pt} \pagestyle{empty}
\begin{center}
\rightline{CCNY-HEP-13/3}
\rightline{July 2013}
\vspace{1truein} {\Large\bfseries
The Isospin Asymmetry in Anomalous Fluid Dynamics }\\
\vskip .1in
\vspace{6pt}
\vskip .1in
{\Large \bfseries  ~}\\
\vskip .1in
{\Large\bfseries ~}\\
{\large\sc D. Capasso, V.P. Nair and J. Tekel}\\
\vskip .2in
{\itshape Physics Department\\
City College of the CUNY\\
New York, NY 10031}\\

\vskip .1in
\begin{tabular}{r l}
E-mail:
&{\fontfamily{cmtt}\fontsize{11pt}{15pt}\selectfont 
dario.capasso@gmail.com}\\
&{\fontfamily{cmtt}\fontsize{11pt}{15pt}\selectfont vpn@sci.ccny.cuny.edu}\\
&{\fontfamily{cmtt}\fontsize{11pt}{15pt}\selectfont jtekel@ccny.cuny.edu; juraj.tekel@gmail.com}\\

\end{tabular}

\fontfamily{cmr}\fontsize{11pt}{15pt}\selectfont
\vspace{.8in}
\centerline{\large\bf Abstract}
\end{center}
The dynamics of fluids in which the constituent particles carry nonabelian charges can be
described succinctly in terms of group-valued variables via a generalization of the co-adjoint orbit action for particles. This formalism, which is particularly suitable for incorporating
anomalies, has previously been used for the chiral magnetic and chiral vorticity effects.
Here we consider the similar effect for the isospin which corresponds to
an angular asymmetry
for neutral pions. 

\end{titlepage}

\pagestyle{plain} \setcounter{page}{2}

Fluid dynamics can be formulated in terms of group-valued variables
\cite{{Bistrovic:2002jx}}.
The basic idea behind this is to construct a Lagrangian generalizing the usual co-adjoint orbit action which describes particle motion as the transport of various conserved quantum numbers, the latter being
the simultaneously diagonalizable generators of  an appropriate group.
Thus, for the transport of mass and spin, one would consider the Poincar\'e group,
or the de Sitter group with a suitable contraction leading to the Poincar\'e group, see
for example \cite{Capasso:2012dr}.
Likewise, the color group, say $SU(3)$, would describe the color flow in the quark-gluon plasma.
One of the advantages of this formulation, as pointed out recently, is that one can incorporate the effect of anomalies very easily via a generalization of the usual Wess-Zumino term
\cite{Nair:2011mk}.

The existence of such an effective action for anomalies is in complete conformity with the 't Hooft argument for 
the usual Wess-Zumino term \cite{'tHooft:1979bh}.
Consider a situation where all the  flavor symmetries of the quarks 
are gauged with anomalies canceled by
a set of spectator fermions. 
Then
 if the quark degrees of freedom are realized in a different phase
 with a different set of variables taking the place of the quark fields, 
there should be a term given in terms of the new variables
which can reproduce the anomalies
 so that the anomaly cancellation with the spectator fermions is still preserved in the new phase.
This is the basic {\it raison d'{\^e}tre} for the Wess-Zumino term.
The original argument by 't Hooft was applied to the phase with confinement and chiral symmetry breaking so that mesons and baryons took the place of the quark degrees of freedom, but 
the same reasoning can be used for the fluid phase, with the quarks replaced by the fluid degrees of freedom such as flavor flow velocities and densities.
The group theoretical formulation of fluid dynamics is especially appropriate for this context, because the original Wess-Zumino action for anomalies is given in terms of the group-valued pseudoscalar meson fields and so, for the fluid phase, essentially the same expression can be used, except for the group-valued fields being reinterpreted in terms of flavor flow velocities and densities.
In this way, one can obtain the chiral magnetic effect (CME) and the chiral vorticity effect from a purely symmetry-based approach. 
We note that beyond the original papers on the CME \cite{Kharzeev:2004ey}, it has been discussed from many different points of view, from a combination of hydrodynamics and thermodynamics
\cite{arXiv:1010.1550}, from holography \cite{arXiv:0904.4772}, using chiral Lagrangians
\cite{fukushima}, from fluid-gravity correspondence
\cite{arXiv:1102.4334}, in different dimensions \cite{loganayagam} and from the lattice
formulaton \cite{arXiv:0907.0494}. The chiral vorticity effect has also been discussed in the literature, see
\cite{landsteiner}. A recent phase space formulation of anomalies is suggestively close to the fluid description we use, although the similarities need further exploration \cite{Stone}.

 {\it A priori}, it may seem that processes mediated by the anomalies cannot be very important since the unconfined fluid phase of quarks and gluons is obtained only for a very short time, of the order of strong interaction scales while flavor processes, typically, are on  a much longer time-scale. 
However, some flavor processes can be enhanced. The high value of the  magnetic field generated by the passing ions in a slightly off-center collision acts as an enhancement factor.
Thus an anomaly mediated process, such as
the electromagnetic current induced via the term responsible for the 
$\pi^0 \rightarrow 2 \, \gamma$, can be significant, because one of the contributing factors is the
magnetic field.
This was the main reason for considering the CME in the context of charge asymmetry in heavy ion collisions. However, we should point out that this enhancement may not be adequate enough to produce observable effects. There are alternate, and arguably compelling, explanations of the charge asymmetry \cite{alternate}, due to correlations imposed by conservation laws, initial state fluctuations, etc.
It could very well be that the observed charge asymmetry  is due to one of these alternate
possibilities, although the relevance of the CME for heavy ion collisions is not entirely a settled matter. Overall, the issue of anomalies in fluid dynamics still
remains a matter of research interest and the present paper is set in this larger context.

The content of this paper can be summarized as follows.
We will explore the fluid action based on group theory in more detail. 
The focus of \cite{Nair:2011mk}, as well as many of the earlier papers, was on one of the anomalies, namely, the fluid version of the term that leads to the standard $\pi^0 \rightarrow 2 \, \gamma$ decay.  However, the standard model has other flavor anomalies and some of these other ones can also have interesting consequences.
The main result of our analysis in this paper is to point out that, similar to the CME, there is also an isospin
asymmetry which can be manifested as an an angular asymmetry in the emission of neutral pions.
There is some enhancement for the latter as well, albeit, not to the same extent as the CME.

We start with a very brief review of the formalism from \cite{{Bistrovic:2002jx},{Nair:2011mk}}.
The key observation is that for the motion of particles which carry a nonabelian
charge corresponding to a Lie group $G$, the action is given by \cite{{wong},{bal1}}
\begin{equation}
S= \int dt \left[ {1\over 2} m \, {\dot x}^2 + A_i^a Q^a {\dot x}_i \right]
\,-i\int dt\sum^r_{s =1}w_{s}\, \Tr\left(q_s g^{-1}\,\dot{g}\right)\label{SM1}
\end{equation}
where we have written the nonrelativistic action for the usual kinetic term in $S$. (It can be made relativistic without affecting the results which follow.)
Further, $g$ is an element of the group $G$, taken to be a matrix in the fundamental representation, with $t^a$ being an orthonormal basis in the same representation,
$\Tr (t^a t^b ) = {1\over 2} \delta^{ab}$. $q_s$ are the diagonal generators in the same basis, i.e., they are the generators of the Cartan subalgebra; thus the range of the summation is from $1$ to the rank of the group, denoted by $r$.
$A_i = -i t^a A_i^a$ is the nonabelian gauge field and
$Q^{a}= \sum_s w_s\,\Tr (g\, q_s \,g^{-1}\,t^{a})$.

The last term in (\ref{SM1}) is the
co-adjoint orbit action which describes the dynamics of the gauge
charges and which, upon quantization, gives the Hilbert space corresponding
to one unitary irreducible representation ($UIR$) of $G$ corresponding
to the highest weight ${\bf w} = \{w_s\}= (w_1, w_2,\cdots, w_r)$.
$Q^{a}$
then become operators realizing the charge algebra
\begin{equation}
\left[Q^{a},Q^{b}\right]=if^{abc}Q^{c}.\label{SM2}
\end{equation}
Under $g\rightarrow g\exp\left(i q_s \,\theta^s\right)$,
the change in the action is given by $\Delta S= {\half}\sum_s w_s \int dt\, {\dot \theta}^s$. One can choose $\theta^s (t)$ such that $\left[ g\,\exp\left(i q_s \,\theta^s\right)\right]$ traces out a closed loop in $G$ over the range of $t$. The single-valuedness
of $e^{iS}$ over such a closed path in $G$ leads
to quantization conditions on ${\bf w}$ corresponding to the UIRs of $G$; consistent quantization is obtained only for the UIRs of $G$.

The generalization of the co-adjoint orbit action for a number of particles would be
\beq
S = -i\int dt\sum_\lambda {\bf w}_\lambda \cdot\, \Tr\left({\bf q}\, g_\lambda^{-1}\,\dot{g}_\lambda\right)
\label{SM3}
\eeq
where $\lambda = 1, 2, \cdots, N$ labels the particles and ${\bf w} \cdot {\bf q}
= \sum_s w_s \, h_s$.
For the fluid viewed as composed of individual particles transporting nonabelian charges,
we may take the continuum limit of this action, as is usually done
in the Lagrange approach to fluids, by replacing the indexing label $\lambda$
by ${\bf x}$, the position of the particle, and with the corresponding changes
$\sum_{\lambda}\rightarrow\int {d^{3}{\bf x}/ v}$, ${w_{\lambda} /v}\rightarrow\rho({\bf x})$.
Here $v$ is a small volume over which the coarse-graining is done to get a continuum description.
This procedure leads to
\begin{equation}
S=-i\int d^{4}x~\sum_s\, \rho_s \, \Tr(q_s\, g^{-1}\, \dot{g})\label{SM4}
\end{equation}
where $g=g({\bf x},t)$. Since the charge density $\rho_s$ can be taken as the time-component of the four-vector current, the relativistic version of (\ref{SM4}) is straightforwardly given by
\beq
S=-i\int d^{4}x~\sum_s\, j_s^{\mu}\, \Tr\left(q_s \,g^{-1}\, D_{\mu}g\right)
\label{SM5}
\eeq
where $D_{\mu}g=\partial_{\mu}g+A_{\mu}g$, $A_{\mu}=-it^{a}A_{\mu}^{a}$.
The inclusion of terms corresponding to the usual terms of the fluid action is also
straightforward. We get
\beq
S=-i \int d^4x\, \sum_{s} \, j_{s}^{\mu}\, \Tr\left(q_{s}g^{-1}D_{\mu}g\right)- \int d^4x~ F(n_{1},n_{2},\ldots)+S_{YM}(A)\label{SM6}
\eeq
where, for each value of $s$, $j_{s}^{\mu}j_{s\mu}=n_{s}^{2}$. The function $F(n_1, n_2, \cdots)$
contains information about the pressure and the enthalpy. We have also introduced the standard Yang-Mills action for the gauge field.

The equations of motion for (\ref{SM6}) do give
the appropriate nonabelian magnetohydrodynamics.
Further, the canonical quantization of this action leads
to the expected current algebra, with the following equal-time
rules for the charge density,
\begin{equation}
\left[\rho^{a}({\bf x},t),\, \rho^{b}({\bf y},t)\right]=if^{abc}\rho^{c}({\bf x},t)\, \delta^{3}({\bf x}-{\bf y}).\label{SM7}
\end{equation}
The velocity for the transport of the nonabelian charge can be introduced
via $j^{\mu}=n\,u^{\mu}$, $u^{2}=1$, for each value of $s$.
Also, from (\ref{SM6}), the current which couples to
the gauge field $A_{\mu}^{a}$ is given by
\begin{equation}
J^{a\mu}=  \sum_s j^\mu_s \, \Tr ( g\, q_s\, g^{-1}\, t^a)
= \sum_s u^\mu_s\, n_s  \, Q^a_s\label{SM8}
\end{equation}
which is in the Eckart form \cite{eckart}, again, for each value of $s$.

We can also interpret the dynamical variables as follows.
The charge density, considered as a matrix in the fundamental representation,
transforms as $\rho\rightarrow h^{-1}\,\rho\, h$, $h\in G$, $\rho=\rho^{a} t^{a}$.
We can thus pick a specific transformation $g$ which diagonalizes
$\rho$,
\begin{equation}
\rho=g\,\rho_{diag}\,g^{-1}, \hskip .3in \rho_{diag} = \sum_s n_s\, q_s
\label{SM9}
\end{equation}
so that $\rho^{a}= \sum_s n_s\,Tr\left(g\, q_s\,g^{-1}\,t^{a}\right)$. Thus $g({\bf x},t)$ 
is part of the charge density and 
$n_s$ are the eigenvalues of $\rho$.
The eigenvalues $n_s$ are gauge-invariant and 
their flow is given by $u_s^{\mu}$.

We can now consider the specialization of the action
(\ref{SM6}) to the
fluid or plasma phase of the standard model. We will consider the quark-gluon
plasma phase for three flavors of quarks, $u,d,s$. In other words,
we consider a phase with thermalized $u,d,s$ quarks, so that they
must be described by fluid variables while the heavier quarks are
described by the field corresponding to each species. 
We will also consider all flavor symmetries to be gauged;
once the action has been written down, we can specialize to the case of the flavor gauge fields being those of the electroweak theory.
We will also neglect the
quark masses so that we have the full flavor symmetry $U(3)_{L}\times U(3)_{R}$.
Thus the group $G$ to be used in (\ref{SM6}) is 
\[
G=SU(3)_{c}\times U(3)_{L}\times U(3)_{R}\label{eq:standard model color and flavor}
\]
with individual flows corresponding to the charges. In this discussion
our focus is on the flavor transport, so we will drop $SU(3)_{c}$
from the equations to follow. Thus the diagonal elements of the algebra are
the $\lambda_3$, $\lambda_8$ and $\lambda_0 = \mathbb{1}$ of
$U(3)_L$ and $U(3)_R$.
The fluid action
is then given by
\begin{eqnarray}
S & = & \int d^4x~\biggl[ -i\, j_{3}^{\mu}\Tr\left(\frac{\lambda_{3}}{2}g_{L}^{-1}D_{\mu}g_{L}\right) -i\, j_{8}^{\mu}\Tr\left(\frac{\lambda_{8}}{2}g_{L}^{-1}D_{\mu}g_{L}\right)
 -i\, j_{0}^{\mu}\Tr\left(g_{L}^{-1}D_{\mu}g_{L}\right)
 \nonumber \\
 & & \hskip .25in -i\,k_{3}^{\mu}\Tr\left(\frac{\lambda_{3}}{2}g_{R}^{-1}D_{\mu}g_{R}\right) - i\,k_{8}^{\mu}\Tr\left(\frac{\lambda_{8}}{2}g_{R}^{-1}D_{\mu}g_{R}\right)
-i\, k_{0}^{\mu}\Tr\left(g_{R}^{-1}D_{\mu}g_{R}\right)\nonumber \\
&& \hskip .25in -F({n_{0},}n_{3},n_{8}, {m_{0},}{m_3},m_{8}) +S_{YM}(A)\nonumber \\
&& \hskip .25in+\,\Gamma_{WZ}(A_{L},A_{R},g_{L}g_{R}^{\dagger}) - 
\Gamma_{WZ}(A_{L},A_{R}, \, {\mathbb 1}) 
\biggr]\label{SM10}
\end{eqnarray}
where $j_{0,3,8}^{\mu}$ apply to $U(3)_{L}$, $k_{0,3,8}^{\mu}$
apply to $U(3)_{R}$, $g_L \in U(3)_L$, $g_R \in U(3)_R$ and $n_{l}^{2}=j_{l}^{2}$, $m_{l}^{2}=k_{l}^{2}$ with $l=0,3,8$.
The last term is the usual gauged WZ term $\Gamma_{WZ}(A_{L},A_{R},U)$
given in terms of $A_{L},A_{R}$ and the meson fields $U\in U(3)$
 \cite{Witten:1983tw,Kaymakcalan:1983qq},
but, for our purpose, $U$ is replaced by $g_{L}g_{R}^{\dagger}$. 
We have also subtracted $\Gamma_{WZ}(A_{L},A_{R}, \, {\mathbb 1}) $ which is necessary to bring the analysis to the Bardeen form of the anomalies
 \cite{Kaymakcalan:1983qq}. Recall that the Bardeen form is the
one which not only preserves the vector gauge symmetries, but also gives a manifestly
vector-gauge-invariant form to the remaining axial anomalies. This form is what is appropriate for the analysis we need. In this sense, the action (\ref{SM10}) is a modification of the action given
in \cite{Nair:2011mk}.
Explicitly $\Gamma_{WZ} (A_L, A_R, U)$
is given by  \cite{Witten:1983tw,Kaymakcalan:1983qq}\footnote{There are some sign differences
with the expression used in \cite{Nair:2011mk}. This removes some sign inconsistencies we had
(which did not affect the results in \cite{Nair:2011mk}) and the present choice is
consistent with \cite{Kaymakcalan:1983qq}.}
\begin{align}
\Gamma_{WZ} (A_l, A_R, U) = & \,C \int \Tr\left(dUU^{-1}\right)^{5}\nonumber\\
 &+5 C\int \Tr\left(A_{L}dA_{L}+dA_{L}A_{L}+A_{L}^{3}\right)dUU^{-1} \nonumber\\
&+5 C\int \Tr\left(A_{R}dA_{R}+dA_{R}A_{R}+A_{R}^{3}\right)U^{-1}dU\nonumber\\
 &- {5 \over 2} C\int \Tr\left[\left(A_{L}dUU^{-1}\right)^{2}-\left(A_{R}U^{-1}dU\right)^{2}\right]\nonumber\\
 & - 5 C\int \Tr\left[A_{L}\left(dUU^{-1}\right)^{3} + A_{R}\left(U^{-1}dU\right)^{3}\right] \nonumber\\
 &- 5 C\int \Tr\left(dA_{L}dUA_{R}U^{-1}-dA_{R}dU^{-1}A_{L}U\right) \nonumber\\
 &-5 C\int \Tr\left(A_{R}U^{-1}A_{L}U(U^{-1}dU)^{2}-A_{L}UA_{R}U^{-1}(dUU^{-1})^{2}\right) \nonumber\\
 &+5 C\int \Tr\left((dA_{R}A_{R}+A_{R}dA_{R})U^{-1}A_{L}U-(dA_{L}A_{L}+A_{L}dA_{L})UA_{R}U^{-1}\right) \nonumber\\
 &+5 C \int \Tr\left(A_{L}UA_{R}U^{-1}A_{L}dUU^{-1}+A_{R}U^{-1}A_{L}UA_{R}U^{-1}dU\right) \nonumber\\
 &+5 C\int \Tr\left(A_{R}^{3}U^{-1}A_{L}U-A_{L}^{3}UA_{R}U^{-1}+\frac{1}{2}UA_{R}U^{-1}A_{L}UA_{R}U^{-1}A_{L}\right)
 \label{SM11}
\end{align}
where $C = - i (N/240\pi^2)$, $N$ being the number of colors ($= 3$ for us).
As observed in \cite{Nair:2011mk} the action given above incorporates all the flavor anomalies in fluid dynamics. It is straightforward to verify that (\ref{SM11}) does indeed lead to the usual chiral magnetic effect. We can now specialize to the problem at hand. We are interested in terms involving the $Z^0$ field and the electromagnetic field. Taking the electromagnetic field to be the magnetic field of the passing ions, which is then the enhancement factor, the $Z^0$-dependent
term can lead to the anomlay-induced weak neutral current.
For this calculation, we may neglect all gauge fields except
$A_\mu$ (the electromagnetic field) and $Z_\mu$.
This means that we can take the left and right gauge fields to be
\beqar
A_{L\mu}
&=& -ieQA_{\mu}
-i\frac{g}{\cos\theta_{W}}(I_{3}-Q\sin^{2}\theta_{W})\,Z_{\mu}\nonumber\\
&=& (-i)\left(\alpha A_{\mu}+\beta Z_{\mu}\right)\label{SM12}\\
A_{R\mu}
&=& -ieQA_{\mu}
-i\frac{g}{\cos\theta_{W}}(-Q\sin^{2}\theta_{W})Z_{\mu}\nonumber\\
&=& (-i)\alpha\left(A_{\mu}-\tan\theta_{W}Z_{\mu}\right)\label{SM13}
\end{eqnarray}
where we have used $e = g \sin\theta_W$ and defined
\beq
\alpha=e\,Q , \hskip .3in
\beta=\frac{g}{\cos\theta_{W}}I_{3}-\alpha\tan\theta_{W}
\label{SM14}
\eeq
The matrices $I_3$ and $Q$ are the usual ones,
\beq
I_3 = \left( \begin{matrix}
{1\over 2}&0&0\\
0&-{1\over 2}&0\\
0&0&-{1\over 2}\\
\end{matrix}\right), \hskip 0.3in
Q = \left( \begin{matrix}
{2\over 3}&0&0\\
0&-{1\over 3}&0\\
0&0&-{1\over 3}\\
\end{matrix}\right)
\label{SM15}
\eeq
Evidently, $Q = I_3 + {\mathbb 1}/6$.
The terms of interest to us are those involving the electromagnetic field and
terms with one power of $Z$. 
The terms in the action (\ref{SM11}) involving just the electromagnetic field are \cite{callan}
\beqar
S^{(1)}_{int} &=&- 5\,C \int \left[ A\, dA \, \Tr \left( 2\,\alpha^2 (dU \,U^{-1} + U^{-1} dU)
- \alpha \,U\, \alpha \, dU^{-1} +  \alpha \, dU\, \alpha \, U^{-1}\right)\right.\nonumber\\
&&\hskip .6in\left. - i A\, \Tr \left( \alpha\, (dU\, U^{-1})^3 + \alpha (U^{-1} dU)^3\right)\right]
\label{SM15a}
\eeqar
These terms lead to the usual chiral magnetic effect as discussed in 
 \cite{Nair:2011mk}. The terms with one power of $Z$
 in $\Gamma_{WZ}$ are collected together as
\beqar
S^{(2)}_{int}
&=&
5\, C\int \Tr\left\{
iZ(\beta-\tan\theta_{W}U\alpha U^{-1})(D_{A}UU^{-1})^{3}
\right. \nonumber\\
&& \left. +ZdA(
-4\alpha\beta
+4\tan\theta_{W}U\alpha^{2}U^{-1}
+\{\tan\theta_{W}\alpha-\beta,U\alpha U^{-1}\}
)D_{A}UU^{-1}
\right\}\label{SM16}
\eeqar
where $D_{A}UU^{-1}$ is the derivative covariant with respect to the electromagnetic field,
\beq
D_{A}UU^{-1}
\equiv dUU^{-1}
-iA(\alpha-U\alpha U^{-1}).
\label{SM17}
\eeq
Explicitly, in local coordinates, the interaction term (\ref{SM16}) is 
\beqar
S_{int}
&\!\!\!=&\!\!\!
-\frac{iN}{48\pi^{2}}\epsilon^{\mu\nu\gamma\delta}\int d^{4}x \,Z_{\mu}\Tr\left\{
i(\beta-\tan\theta_{W}U\alpha U^{-1})(D_{A}UU^{-1})_{\nu}(D_{A}UU^{-1})_{\gamma}(D_{A}UU^{-1})_{\delta}
\right.\nonumber\\
&&\!\! \left. +\, {\del_\nu A_\gamma}(
-4\alpha\beta
+4\tan\theta_{W}U\alpha^{2}U^{-1}
+\{\tan\theta_{W}\alpha-\beta,U\alpha U^{-1}\}
)(D_{A}UU^{-1})_{\delta}
\right\}
\label{SM18}
\eeqar
It is useful to consider the reduction of this expression for the case of two flavors,
as this is adequate for illustrating the result on neutral currents. For this, we take
\beq
U = e^{i \theta} \left( \begin{matrix}
V&0\\  0&1\\ \end{matrix}\right)
\label{SM19}
\eeq
where $V$ is a $2\times 2$ $SU(2)$ matrix. We may take it to be of the form
$h_L h_R^\dagger$ where $h_L$ and $h_R$ are elements of $SU(2)$. They can be related to the flow velocities by using the equations of motion for the full action
(\ref{SM10}).
The derivatives now simplify as
\begin{align}
dU \, U^{-1} = i \, d\theta \, {\mathbb 1} + \left( \begin{matrix}
dV\, V^{-1} &0\\  0& 0\\ \end{matrix} \right), &
\hskip .1in
D_{A}U \, U^{-1} = i \, d\theta \, {\mathbb 1} + \left( \begin{matrix}
D_{A} V\, V^{-1} &0\\  0& 0\\ \end{matrix} \right)\nonumber\\
D_{A} V\, V^{-1} = dV \, V^{-1} - & i e A \left( {\tau_3 \over 2} - V {\tau_3\over 2} V^{-1}
\right)
\label{SM20}
\end{align}
where $\tau_3$ is the diagonal Pauli matrix.
The $A$-dependent terms in (\ref{SM15a}) then simplify as
\begin{align}
S^{(1)}_{int} &= - 20\, C i \, e^2 \int A \, dA\, d\theta \nonumber\\
&\hskip .1in - {5\over 2} C i\,e^2 \, \int\Tr \bigl[ \tau_3 V \tau_3 V^{-1} - {\mathbb 1}\bigr] \, A \, dA\, d\theta
- {5 \over 2} C e^2\,\int A\, dA\, \Tr \bigl[ \tau_3 (dV\, V^{-1} + V^{-1} dV )\bigr]\nonumber\\
&\hskip .1in - {5 \over 2} C e\int \,A\,d\theta\, \Tr \bigl[ \tau_3 (dV\,V^{-1})^2 + \tau_3 (V^{-1} dV)^2\bigr]
\nonumber\\
&\hskip .1in + {5\over 2} C i\, e\,\int A\, \Tr \bigl[ \tau_3 (dV \, V^{-1})^{3} + \tau_3 (V^{-1} dV)^{3}\bigr]
+ {5\over 3} C i \, e\, \int A\, \Tr \bigl[ (dV \, V^{-1})^3\bigr]
\label{SM21}
\end{align}
If we set $h_L = h_R$ or $V= {\mathbb 1}$, which is adequate for the basic chiral magnetic effect,
\beq
S^{(1)}_{int} = - 20\, C\, i e^2\, \int A\, dA\, d\theta
\label{SM22}
\eeq
The electromagnetic current which follows from this is
\begin{align}
J^{em} & = - {N e^2 \over 6 \pi^2} \, dA\, d\theta\nonumber\\
J^{em\, \mu} &= - {e^2 \over 4\pi^2} \epsilon^{\mu\nu\alpha\beta}\, F_{\nu\alpha}
\del_\beta \theta
\label{SM23}
\end{align}
where the second line gives the component-form for $N =3$. We can do a similar simplification of the terms involving $Z$ to get
\begin{align}
S^{(2)}_{int} &= 5\,Ci\, (- 8 e^2 \cot2\theta_W )\int Z\, dA\, d\theta\nonumber\\
&\hskip .1in
- 5 Ci e^2 \cot2\theta_W\, \int \Tr (\tau_3 V \tau_3 V^{-1} -{\mathbb 1})\, Z\, dA\, d\theta\nonumber\\
&\hskip .1in - {5\over 2}C e \int Z\, d\theta \left[
 \cot\theta_W\, \Tr \bigl[\tau_3 (DV\, V^{-1} )^2\bigr] -  \tan\theta_W \, \Tr \bigl[\tau_3 (V^{-1} DV)^2\bigr]\right]\nonumber\\
&\hskip .1in
 + {5\over 2} Ce^2 \int Z\, dA\, \left[ \left(
-\cot\theta_W+\tan\theta_W \right) \,\Tr (\tau_3 DV\, V^{-1})
+2\tan\theta_W \, \Tr (\tau_3 V^{-1} DV)\right]\nonumber\\
&\hskip .1in + {5\over 2}C i e \int Z\left[ \cot\theta_W \Tr \bigl[ \tau_3 (DV\,V^{-1})^3\bigr]
- \tan\theta_W \Tr \bigl[ \tau_3 (V^{-1}DV)^3\bigr] \right]\nonumber\\
&\hskip .1in - {5\over 3}C i e \tan\theta_W \int Z\,  \Tr (DV\, V^{-1})^3
\label{SM24}
\end{align}
Again, if we set $V = {\mathbb 1}$, this reduces to
\beq
S^{(2)}_{int} = - {Ne^2 \over 6 \pi^2} (\cot 2\theta_W)\, \int Z\, dA\, d\theta
\label{SM25}
\eeq
The standard coupling of $Z$ to the neutral current is of the form $(g /\cos\theta_W) Z_\mu J^{Z\, \mu}$, so that, we can identify the anomaly-induced neutral current as
\beqar
J^Z &=& - {Ne \over 12\pi^2} (\cos 2\theta_W)\, dA\, d\theta\nonumber\\
J^{Z\, \mu} &=& - {e \over 8\pi^2} (\cos 2\theta_W)\, \epsilon^{\mu\nu\alpha\beta}
F_{\nu\alpha} \del_\beta \theta
\label{SM26}
\eeqar
Since the current for the third component of the weak isospin is
$J^3 = J^Z + \sin^2\theta_W \, (J^{em}/e)$, we get
\beq
J^{3\, \mu} = - {e \over 8\pi^2}  \, \epsilon^{\mu\nu\alpha\beta}
F_{\nu\alpha} \del_\beta \theta
\label{SM27}
\eeq
In the quark-gluon fluid, we may replace ${\dot\theta}$ in terms of the chemical potentials for the left and right axial charges as by ${\dot \theta} = {\half}(\mu_L - \mu_R)$, and so the
spatial component of this current can be written as
\beq
J^{3\, i} = {e \over 8 \pi^2} (\mu_L - \mu_R)\,B^i
\label{SM28}
\eeq
Thus there is an induced weak isospin asymmetry possible.

We now turn to the interpretation of this induced current.
In the region where we have the quark-gluon fluid, there is a current in the direction of the magnetic field. This corresponds to a flow of the constituent particles of the medium. Continuity
requires that such a current should exist just outside of the fluid region where the degrees of freedom are the hadrons. In other words, we expect this current to translate into 
the hadronic version of the weak isospin current just outside of the fluid region.
Since pions are the most significant component of the hadrons, the current of interest is
$J^{3\,\mu} = - {\half}{f_\pi} \, \del^\mu \Pi^0 + \cdots$.
This shows that if we consider $\Pi^0$ as a classical field configuration, then the result (\ref{SM27}) can be interpreted as saying that a gradient in the pion field is generated by the anomaly; it is given by
\begin{align}
\del^\mu \Pi^0 &= {e \over 4 \pi^2  f_\pi} \, \epsilon^{\mu\nu\alpha\beta} F_{\nu\alpha} \del_\beta \theta\nonumber\\
\del^i \Pi^0 &= - {e \over 4 \pi^2  f_\pi} \,(\mu_L - \mu_R) \, B^i
\label{SM29}
\end{align}
Effectively, this implies an asymmetry in the distribution of neutral pions, in the direction of the magnetic field of the passing heavy nuclei. More explicitly, since pions are detected via the $2 \, \gamma$ final states, consider the effective Lagrangian for this decay,
\beq
S_{eff} = - {\alpha_e \over 4 \pi f_\pi} \int d^4x~ \epsilon^{\mu\nu\alpha\beta}
\del_\mu \Pi^0 \, F_{\nu \alpha} A_\beta
\label{SM30}
\eeq
This shows that a classical pion field may be thought of as an antenna for the radiation of
correlated photons. If we use
(\ref{SM29}), we may even write (\ref{SM30}) as
\begin{align}
S_{eff} & = - {\alpha_e e\over 8 \pi^3 f_\pi^2} \int \left[
{\bar F}^{\mu\nu} \, F_{\mu\nu} \,A_\alpha \del^\alpha \theta
+ 2\,{\bar F}^{\mu\nu} F_{\alpha \mu} A_\nu \del^\alpha\theta
\right]\nonumber\\
&= - {\alpha_e e\over 8 \pi^3 f_\pi^2} (\mu_L - \mu_R) \int 
\epsilon^{ijk} {\bar B}_i \, F_{0j} A_k
\label{SM31}
\end{align}
where we have indicated the magnetic field generated by the passing ions with an overbar; the other two fields correspond to the radiated photons.
The magnitude of this effect remains small, of the order of the CME, with a further
suppression due to the pion decay, via $f_\pi$ in (\ref{SM29}). (Here we are not counting
the additional
factor of $\alpha_e /f_\pi$ due to (\ref{SM30}), since it is there for any observed pion decay.)

Another effect of the action (\ref{SM25}) is to consider it as an interaction term generating a virtual $Z^0$ which can then decay into leptons. Effectively, this amounts to replacing
$Z_\mu$ by $(g/\cos\theta_W) (J^Z/M_Z^2)$, so that
\beqar
S^{(2)}_{int} &\approx& - {Ne^2 \over 12\pi^2} (\cot2\theta_W) {g \over M_Z^2\, \cos\theta_W}
\int d^4x~ J^Z_\mu F_{\nu\alpha} \del_\beta \theta \, \epsilon^{\mu\nu\alpha\beta}\nonumber\\
&\approx& {Ne^2 g \over 12\pi^2} {(\cot2\theta_W)  \over M_Z^2\,\cos\theta_W}
(\mu_L - \mu_R)\,\int d^4x~ J^Z_i {\bar B}_i 
\label{SM32}\\
J^Z_i &=& {1\over 2} \left( {\bar \nu}_{e L} \gamma_i \nu_e - {\bar e}_L \gamma_i e_L
\right) + \sin^2\theta_W ({\bar e}\gamma_i e ) + \cdots
\nonumber
\eeqar
This can show up as an asymmetry in the lepton distribution, particularly for neutrinos, although experimentally it would be too small for detection. 
 Part of this effect will also act as a modification of the charged particle asymmetry.
The effect is much smaller than the chiral magnetic effect, because of the suppression by
$M_Z^{-2}$. With more energetic collisions, as at the LHC, the transient magnetic field is higher,
which gives somewhat better enhancement; one can also get some kinematic enhancement from the current for highly energetic
leptons. These can mitigate the effect of $M_Z^{-2}$ to some extent.

\vskip .1in
VPN and JT were supported in part by NSF grant PHY-1213380 
and by PSC-CUNY grants. DC was supported by a Templeton Foundation grant
21531.

\end{document}